\documentclass[11pt,a4paper]{article}
\usepackage{jcappub}

\usepackage[latin1]{inputenc}
\usepackage{graphicx}
\usepackage{epsfig}
\usepackage{color}
\usepackage{comment}
\usepackage{slashed}

\title{Multimessenger tests of the weak equivalence principle from GW170817 and its electromagnetic counterparts}

\author[a]{Jun-Jie Wei,}
\author[b]{Bin-Bin Zhang,}
\author[a,c]{Xue-Feng Wu,}
\author[d]{He Gao,}
\author[e,f,g]{Peter M{\'e}sz{\'a}ros,}
\author[h]{Bing Zhang,}
\author[i]{Zi-Gao Dai,}
\author[j,k]{Shuang-Nan Zhang}
\author[d]{and Zong-Hong Zhu}

\affiliation[a]{Purple Mountain Observatory, Chinese Academy of Sciences, Nanjing 210008, China}
\affiliation[b]{Instituto de Astrof\'isica de Andaluc\'a (IAA-CSIC), P.O. Box 03004, E-18080 Granada, Spain}
\affiliation[c]{Joint Center for Particle, Nuclear Physics and Cosmology, Nanjing University-Purple Mountain Observatory, Nanjing 210008, China}
\affiliation[d]{Department of Astronomy, Beijing Normal University, Beijing 100875, China}
\affiliation[e]{Department of Astronomy and Astrophysics, Pennsylvania State University, 525 Davey Laboratory, University Park, Pennsylvania 16802, USA}
\affiliation[f]{Department of Physics, Pennsylvania State University, 104 Davey Laboratory, University Park, Pennsylvania 16802, USA}
\affiliation[g]{Center for Particle and Gravitational Astrophysics, Institute for Gravitation and the Cosmos, Pennsylvania State University, 525 Davey
Laboratory, University Park, Pennsylvania 16802, USA}
\affiliation[h]{Department of Physics and Astronomy, University of Nevada Las Vegas, Las Vegas, Nevada 89154, USA}
\affiliation[i]{School of Astronomy and Space Science, Nanjing University, Nanjing 210093, China}
\affiliation[j]{Laboratory for Particle Astrophysics, Institute of High Energy Physics, Beijing 100049, China}
\affiliation[k]{National Astronomical Observatories, Chinese Academy of Sciences, Beijing 100012, China}

\emailAdd{xfwu@pmo.ac.cn}

\abstract{The coincident detection of a gravitational-wave (GW) event GW170817 with electromagnetic (EM) signals
(e.g., a short gamma-ray burst SGRB 170817A or a macronova) from a binary neutron star merger within
the nearby galaxy NGC 4933 provides a new, multimessenger test of the weak equivalence principle (WEP),
extending the WEP test with GWs and photons. Assuming that the arrival time delay between the GW signals
from GW170817 and the photons from SGRB 170817A or the macronova is mainly attributed to the gravitational
potential of the Milky Way, we demonstrate that the strict upper limits on the deviation from the WEP
are $\Delta \gamma<1.4\times10^{-3}$ for GW170817/macronova and $\Delta \gamma <5.9\times10^{-8}$
for GW170817/SGRB 170817A. A much more severe constraint on the WEP accuracy can be achieved
($\sim0.9\times10^{-10}$) for GW170817/SGRB 170817A when we consider the gravitational potential
of the Virgo Cluster, rather than the Milky Way's gravity. This provides the tightest limit to date
on the WEP through the relative differential variations of the $\gamma$ parameter for two different
species of particles. Compared with other multimessenger (photons and neutrinos) results, our limit
is 7 orders of magnitude tighter than that placed by the neutrinos and photons from supernova 1987A,
and is almost as good as or is an improvement of 6 orders of magnitude over the limits obtained by
the low-significance neutrinos correlated with GRBs and a blazar flare.}


\arxivnumber{1710.05860}

\begin{document}
\maketitle

 \flushbottom


\section{Introduction}
The Laser Interferometer Gravitational-Wave Observatory (LIGO) has previously identified four gravitational
wave (GW) sources with high statistical significance, GW150914 \cite{2016PhRvL.116f1102A}, GW151226 \cite{2016PhRvL.116x1103A},
GW170104 \cite{2017PhRvL.118v1101A}, and GW170814 \cite{2017arXiv170909660T}, as well as a less significant candidate LVT151012
\cite{2016PhRvD..93l2003A}. All of them are produced by the collisions of binary black holes (BHs). No statistically
significant electromagnetic (EM) counterparts of these GW signals have so far been confirmed by follow-up
observations. Note that a suspected association between GW150914 and a weak gamma-ray transient was
reported by the Fermi Gamma-Ray Monitor (GBM) team \cite{2016ApJ...826L...6C}, despite the lack of
a common consensus for a production mechanism of such counterpart to a BH-BH merger \cite{2016ApJ...819L..21L,2016ApJ...821L..18P,2016ApJ...827L..31Z,2017ApJ...839L...7D}.
Besides BH-BH mergers, it was taken to be just a matter of time before GW signals from binary neutron stars
(NSs) and/or BH-NS mergers would be detected as well \cite{2010CQGra..27q3001A}. Merging BH-NS and NS-NS are not
only important GW sources, but also promising candidates for coincident EM counterparts. These systems are
expected to be accompanied by a variety of detectable EM transients, including short Gamma-Ray Bursts (SGRBs)
\cite{1986ApJ...308L..43P,1986ApJ...308L..47G,1989Natur.340..126E,1992ApJ...395L..83N}, on-beam SGRB afterglows
\cite{1997ApJ...476..232M}, and macronovae/kilonovae \cite{1998ApJ...507L..59L,2012ApJ...746...48M,2013ApJ...774L..23B,2016ARNPS..66...23F}.

On 17 August 2017 at $12:41:04$ UTC, a new kind of gravitational-wave event (GW170817) from a merger of binary NSs
was discovered with high significance by the two Advanced LIGO detectors and the Advanced Virgo detector \cite{2017arXiv171005832T}.
A network of three detectors significantly improves the sky localization of GW170814, resulting in a small probability region
of only 28 $\rm deg^{2}$ (90\% credibility). This source had a luminosity distance of $40^{+8}_{-14}$ Mpc.
Meanwhile, a short Gamma-Ray Burst (SGRB 170817A) triggered GBM on board the Fermi satellite at $T_{0}=12:41:06$ UTC,
about 1.7 s after the merger time \cite{2017GCN.21520....1V,2017arXiv171005446G,2017arXiv171005834L,2017arXiv171005851Z}.
SGRB 170817A was also detected by the International Gamma-Ray
Astrophysics Laboratory (INTEGRAL) \cite{2017arXiv171005449S}. The temporal and spatial coincidence between
GW170817 and SGRB 170817A appears to confirm the long-held hypothesis that NS-NS mergers are linked to SGRBs.
Even more encouragingly, optical follow-ups of the sky localization of GW170817 identified a bright optical counterpart
(SSS17a, now with the IAU identification of AT2017gfo) in NGC 4993 (at 40 Mpc) less than 11 hours
after the merger time, consistent with the localization and distance inferred from gravitational-wave data \cite{2017arXiv171005452C}.
The same optical transient was independently confirmed by several teams \cite{2017arXiv171005833L,2017arXiv171005459S,2017arXiv171005854V,2017arXiv171005843A,2017arXiv171005455T,2017arXiv171005461L,2017arXiv171005462H},
which is believed to be powered by
the radioactive decay of heavy elements formed by neutron capture called a macronova/kilonova.
These multimessenger data provide the first firm evidence that GW170817, SGRB 170817A,
and macronova originate from the NS-NS merger.
NGC 4993, the host galaxy of the optical transient, is an elliptical galaxy in the constellation Hydra,
with coordinates (J2000) R.A.=$13^{\rm h}09^{\rm m}47^{\rm s}.7$ and Dec.=$-23^{\circ}23^{'}02^{''}$ \cite{2006AJ....131.1163S}.
In this work, we adopt the more precise location of NGC 4993 as the locations of SGRB 170817A and the
associated macronova, and take its distance as $d=40$ Mpc (corresponding to a redshift of $z=0.009787$)
for the SGRB 170817A and the macronova accordingly.

With the physical association between gravitational and EM waves, the flight time differences between these multimessenger signals
can in principle be used to give important constraints on the validity of Einstein's weak equivalence principle (WEP). The WEP is the foundation
of general relativity as well as of many other metric theories of gravity \cite{2006LRR.....9....3W,2014LRR....17....4W}. In the parametrized post-Newtonian (PPN)
formalism, each gravity theory incorporating the WEP is specified by a set of PPN parameters (e.g., the parameter $\gamma$
which denotes the level of space curvature by unit rest mass), which predicts that the parameter $\gamma$ should be the same
for any two different species of massless (or negligible rest mass) neutral particles, or any two particles of the same species
with different frequencies (i.e., $\gamma_{1}=\gamma_{2}\equiv\gamma$, where the subscripts stand for two different particles) \cite{2006LRR.....9....3W,2014LRR....17....4W}.
Hence, the WEP accuracy can be described by constraints on the differences of the $\gamma$ values for different particles.

In the multimessenger era, the arrival time delays of the same species of particles (e.g., photons, neutrinos, or GWs)
but with varying energies have been widely adopted to test the WEP through the relative differential variations of the $\gamma$
values,\footnote{Note that the arrival time delays of photons with different polarizations from astrophysical events have also
been used to constrain the differences of the $\gamma$ values \cite{2017PhRvD..95j3004W,2017MNRAS.469L..36Y}.},
such as the particle emissions from supernova SN 1987A \cite{1988PhRvL..60..173L}, GRBs \cite{2015ApJ...810..121G,2016MNRAS.460.2282S},
fast radio bursts \cite{2015PhRvL.115z1101W,2016ApJ...820L..31T}, TeV blazars \cite{2016ApJ...818L...2W},
the Crab pulsar \cite{2016PhRvD..94j1501Y,2017ApJ...837..134Z,2016arXiv161202532D}, and GW sources \cite{2016PhRvD..94b4061W,2016PhLB..756..265K,2016ApJ...827...75L}.
However, there are only a few WEP tests with different species of messenger particles, and moreover the tests
have been limited so far to the photon and neutrino sectors. For example, the flight time differences between
MeV neutrinos and photons from SN 1987A have been used to test the WEP accuracy, and the results showed that
the $\gamma$ value for photons is the same as that for neutrinos to within $0.2-0.5\%$ \cite{1988PhRvL..60..173L,1988PhRvL..60..176K}.
With the assumption that the arrival time delay between a PeV neutrino and gamma-ray photons from a blazar
flare is caused dominantly by the gravitational potential of Virgo Cluster, ref. \cite{2016PhRvL.116o1101W} set a
stricter limit on the $\gamma$ differences of $\Delta \gamma<3.4\times10^{-4}$. Based on the associations
between the TeV neutrinos and gamma-ray photons from five GRBs, ref. \cite{2016JCAP...08..031W} tightened the constraint
on the WEP to $\Delta \gamma \sim 10^{-11}-10^{-13}$ when taking into account the gravitational potential of
the Laniakea supercluster of galaxies.  However, except for the SN neutrinos with SN 1987A, the significance
of these TeV neutrinos (or the PeV neutrino) being associated with GRBs (or the blazar flare) is low.
A 5\% probability for a chance coincidence between the blazar flare and the PeV neutrino remains
\cite{2016NatPh..12..807K}, and the coincidences between the TeV neutrinos and GRBs only yielded a combined
p-value of 0.32 \cite{2016ApJ...824..115A}. In sum, except for the WEP test from SN neutrinos, all the other
multimessenger tests have relied on using low-significance neutrino events correlated with photons,
which are not very reliable.
New multimessenger signals exploiting different emission channels with high significance (e.g., GWs and
photons/neurtinos) are essential for further testing the WEP to a higher accuracy level.

Recently, ref. \cite{2016PhRvD..94b4061W} proposed that more tests of the WEP between GWs and photons would be possible
with future coincident detection of GWs with EM counterparts from binary NSs and NS-BH mergers, and they
provided estimates of the resulting constraints on the WEP from various detectable EM counterparts. Based
on the first claimed EM counterpart of GW150914 \cite{2016ApJ...826L...6C}, ref. \cite{2016PhRvD..94b4061W} also showed how
this GW/EM association could be used to test the WEP (see also \cite{2017PhLB..770....8L}). However, BH-BH mergers are
not expected to have enough surrounding material to power bright EM counterparts \cite{2016ApJ...819L..21L,2016ApJ...821L..18P,2016ApJ...827L..31Z,2017ApJ...839L...7D}, it is
still unclear whether the sub-threshold transient event was associated with GW150914 or a chance coincidence.
Our interest in the possibility of using GWs to test the WEP has recently been revived by the observations
of a new kind of gravitational-wave signal (GW170817) which comes from a binary NSs merger. Since GW170817
is confirmed to be accompanied by SGRB 170817A and a macronova with high significance, we now use this true
GW/EM association (i.e., GW170817/SGRB 170817A and GW170817/macronova) to provide robust constraints on the
WEP, thus extending significantly the WEP test with GWs and photons.

\section{WEP tests with GW170817 and its EM counterparts}

In the presence of a gravitational potential $U(r)$, the time interval required for particles to traverse
a given distance is expected to be longer, by the so-called Shapiro (gravitational) time delay \cite{1964PhRvL..13..789S},
\begin{equation}
t_{\rm gra}=-\frac{1+\gamma}{c^3}\int_{r_e}^{r_o}~U(r)dr\;,
\end{equation}
where the integration is along the path of the particle emitted at $r_e$ and observed at $r_o$.
If the WEP fails, the $\gamma$ values of different particles will be different, leading to arrival-time
differences of the two particles arising from the same transient source.
The relative Shapiro time delay is therefore expressed as
\begin{equation}
\Delta t_{\rm gra}=\frac{\gamma_{\rm 1}-\gamma_{\rm 2}}{c^3}\int_{r_e}^{r_o}~U(r)dr\;,
\label{gra}
\end{equation}
where the difference of the $\gamma$ values $\Delta\gamma=\gamma_{\rm 1}-\gamma_{\rm 2}$ can be characterized
as a measure of a possible deviation from the WEP.

To calculate the relative Shapiro time delay with Eq.~(\ref{gra}), we need to figure out the gravitational
potential $U(r)$ along the propagation path. In principle, $U(r)$ consists of three parts: the gravitational
potentials of the Milky Way $U_{\rm MW}(r)$, the intergalactic space $U_{\rm IG}(r)$, and the transient host
galaxy $U_{\rm host}(r)$. Since the potential function of $U_{\rm host}(r)$ and $U_{\rm IG}(r)$ is hard to
model, we first consider only the potential of the Milky Way $U_{\rm MW}(r)$.  Assuming that the observed
time delay ($\Delta t_{\rm obs}$) between correlated particles is dominated by the relative Shapiro time delay,
and adopting a Keplerian potential $U_{\rm MW}(r)=-GM/r$ for the Milky Way,\footnote{Although the gravitational potential
of our galaxy is still unknown, ref.~\cite{1988PhRvL..60..176K} examined two popular potential models, i.e.,
the Keplerian potential and the isothermal potential. They suggested that different potential models do not
have a strong influence on the resulting constraints on the WEP. Here we adopt the Keplerian potential for the Milky Way.}
a conservative upper limit on $\Delta\gamma$ can be estimated to be \cite{1988PhRvL..60..173L,2016PhRvD..94b4061W}
\begin{equation}\label{gammadiff}
\begin{aligned}
\Delta t_{\rm obs} &>\Delta t_{\rm gra}= \Delta\gamma \frac{GM_{c}}{c^{3}} \times \\
          &\ln \left\{ \frac{ \left[d+\left(d^{2}-b^{2}\right)^{1/2}\right] \left[r_{c}+\left(r_{c}^{2}-b^{2}\right)^{1/2}\right] }{b^{2}} \right\}\;,
\end{aligned}
\end{equation}
where $M_{c}\simeq6\times10^{11}M_{\odot}$ is the total mass of the gravitational field source (the Milky Way)
\cite{2011MNRAS.414.2446M,2012ApJ...761...98K}, $d$ is the distance from the particle transient (GW170817) to the center of the
gravitational source (for an extragalactic or cosmic transient, $d$ is approximated as the distance from the
Earth to the particle transient), $r_{c}\simeq8.3$ kpc represents the distance from the Milky Way center
to Earth, and $b$ corresponds to the impact parameter (see Fig. 1 in ref. \cite{1988PhRvL..60..173L}).
The location of the Milky Way center is R.A.=$17^{\rm h}45^{\rm m}40^{\rm s}.04$ and
Dec.=$-29^{\circ}00^{'}28^{''}.1$ \cite{2009ApJ...692.1075G}.

Note that \emph{Insight}-HXMT did not detect any significant gamma-rays between 200 keV and 5 MeV during the SGRB 170817A episode \cite{2017GCN.21518....1V}.
Upper limits for SGRB 170817A are reported in ref. \cite{2017arXiv171006065L}. However, Insight-HXMT was monitoring GW170817 before and after the GW event,
and thus excludes a classical short/hard GRB during this period (upper limits also given) \cite{2017arXiv171006065L}.
Consequently, it is justified to use the time-delay of 1.7 s between GW170817 and SGRB 170817A to test the WEP.
Since the first detection of the optical counterpart (macronova) was carried out in less than 11 hours after the merger time \cite{2017arXiv171005452C},
we adopt 11 hours as the conservative limit of the observed time delay between GW170817 and macronova.
With these time delays and considering
the gravitational potential of the Milky Way, a stringent limit on the WEP from Eq.~(\ref{gammadiff}) is
\begin{equation}
|\gamma_{g}-\gamma_{\gamma}|<1.4\times10^{-3}
\end{equation}
for GW170817/macronova, and
\begin{equation}
|\gamma_{g}-\gamma_{\gamma}|<5.9\times10^{-8}
\end{equation}
for GW170817/SGRB 170817A.\footnote{Two independent works were simultaneously carried out
by refs. \cite{2017arXiv171005834L,2017arXiv171005805W}, who tested the WEP by using only
the observed time delay between GW170817 and SGRB 170817A, and by considering only the effect
of the Milky Way's gravity.}

On the other hand, it has been shown that considering the large-scale gravitational potential in the
intergalactic space would improve the constraints on the WEP accuracy by a few orders of magnitude,
rather than the Milky Way's gravity \cite{2016ApJ...821L...2N,2016arXiv160104558Z,2016JHEAp...9...35L}.
That is, for the cosmic transients, the Shapiro delay due
to nearby clusters and/or superclusters is more important than the Milky Way and the transient host galaxy.
Thus, we here also consider the gravitational potential of the Virgo Cluster. The Virgo Cluster is the
nearest cluster to the Milky Way, whose center lies at a distance of $r_{c}\simeq16.5$ Mpc
at a position of R.A.=$186.8^{\circ}$ and Dec.=$12.7^{\circ}$ (or R.A.=$12^{\rm h}27^{\rm m}12^{\rm s}$
and Dec.=$12^{\circ}42^{'}00^{''}$) \cite{2007ApJ...655..144M}. Its mass is estimated to be
$1.2\times10^{15}M_{\odot}$ out to 8 degrees from the center of the cluster or a radius of about 2.2 Mpc
\cite{2001A&A...375..770F}.  Since the distance of GW170817 is far beyond the scale of the Virgo Cluster, the
gravitational potential of the particle paths from GW170817 to us can be treated as a point mass potential
for which the Virgo Cluster's total mass is assumed at the center of the mass.  Considering the Virgo
Cluster's gravity, a stronger limit on the WEP from Eq.~(\ref{gammadiff}) is
\begin{equation}
|\gamma_{g}-\gamma_{\gamma}|<2.1\times10^{-6}
\end{equation}
for GW170817/macronova, and
\begin{equation}
|\gamma_{g}-\gamma_{\gamma}|<9.2\times10^{-11}
\end{equation}
for GW170817/SGRB 170817A, which is about 6 orders of magnitude tighter than the previous limit obtained
with the PeV neutrino from a blazar flare
($|\gamma_{\nu}-\gamma_{\gamma}|<3.4\times10^{-4}$) \cite{2016PhRvL.116o1101W}, and is almost as good as the results
with TeV neutrinos from GRBs ($|\gamma_{\nu}-\gamma_{\gamma}|<10^{-11}-10^{-13}$) \cite{2016JCAP...08..031W}.

\section{Summary and discussion}

Very recently, the LIGO-Virgo team reported the first detection of a new kind of gravitational-wave event
(GW170817), which arises from a binary NSs merger \cite{2017arXiv171005832T}. Moreover, the EM counterparts of
GW170817, such as the SGRB 170817A and a macronova within the nearby galaxy NGC 4933, are firmly detected
by the follow-up observations \cite{2017GCN.21520....1V,2017arXiv171005446G,2017arXiv171005834L,2017arXiv171005851Z,2017arXiv171005449S,
2017arXiv171005452C,2017arXiv171005833L,2017arXiv171005459S,2017arXiv171005854V,2017arXiv171005843A,2017arXiv171005455T,2017arXiv171005461L,2017arXiv171005462H}.
This is the first time in history that
gravitational and EM waves from a single astrophysical source have been identified. Based on this first
truly GW/EM association, we demonstrate that new, multimessenger WEP tests can be carried out by using
the arrival time delays between the GW signals from GW170817 and the photons from SGRB 170817A or the
macronova. Attributing the time delays to the gravitational potential of the Milky Way, we set stringent
limits on the differences of the PPN $\gamma$ parameter values for two cases, i.e.,
$\Delta \gamma<1.4\times10^{-3}$ for the GW170817/macronova, and $\Delta \gamma <5.9\times10^{-8}$ for
the GW170817/SGRB 170817A. If the time delays are mainly caused by the gravitational potential of the
Virgo Cluster, much more severe constraints can be achieved, implying $\Delta \gamma<2.1\times10^{-6}$
for GW170817/macronova, and $\Delta \gamma<9.2\times10^{-11}$ for GW170817/SGRB 170817A.

To date, only three WEP tests with different species of messenger particles have been obtained, and
moreover these tests were limited to the photon and neutrino sectors. Note that, except for the WEP
test from SN neutrinos \cite{1988PhRvL..60..173L,1988PhRvL..60..176K}, the other two multimessenger tests have relied on
using the low-significance neutrino events correlated with photons \cite{2016PhRvL.116o1101W,2016JCAP...08..031W}, which are
not very reliable. It is highly desirable to develop more accurate tests that include new multimessenger
signals with higher significance (e.g., GWs and photons/neutrinos). In the present work, we have extended
the particle sector over which the WEP is tested to the GWs and photons, at comparable or higher levels
of accuracy. The results from GW170817/SGRB 170817A that we discussed here provide the hitherto most
stringent constraint on the WEP through the relative differential variations of the $\gamma$ parameter
for two different species of particles, namely $\sim10^{-10}$, which is 7 orders of magnitudes tighter
than that placed by the neutrinos and photons firmly detected from SN 1987A \cite{1988PhRvL..60..173L,1988PhRvL..60..176K},
and is as good as or is an improvement of 6 orders of magnitude over the results set by the low-significance
neutrinos correlated with GRBs and a blazar flare \cite{2016PhRvL.116o1101W,2016JCAP...08..031W}. In conclusion, the coincident
detection of GWs with EM signals from compact binary mergers containing NSs can provide attractive
candidates for constraining the WEP, extending the tests to the GW and photon sectors. If in the future
more robust GW/EM associations are detected by the GW detectors and the conventional EM telescopes,
better limits on the accuracy of the WEP might be achieved.

The LIGO detection of GWs from binary BH mergers triggered a large number of follow-up searches for
coincident EM emission \cite{2016ApJ...826L..13A} as well as coincident neutrinos \cite{2016PhRvD..93l2010A,2016PhRvD..94l2007A,2016ApJ...830L..11A},
but no neutrino emission was found. Since in general little or no matter is expected to be present in
the binary BH environment, no significant neutrino emission is expected from such mergers.
However, it will be possible to detect neutrino emission from other source classes, such as NS-NS
and BH-NS mergers, where one does expect neutrino emission, e.g. \cite{2017ApJ...848L...4K}. In view of the
onset of the multimessenger era, a neutrino detection of a GW source would shine a unique light on
testing the WEP with GWs and neutrinos.

\acknowledgments
This work is partially supported by the National Basic Research Program (``973'' Program)
of China (Grant No. 2014CB845800), the National Natural Science Foundation
of China (Grants Nos. 11433009, 11673068, 11603076, and 11725314), the Youth Innovation
Promotion Association (2011231 and 2017366), the Key Research Program of Frontier Sciences (QYZDB-SSW-SYS005),
the Strategic Priority Research Program ``Multi-waveband gravitational wave Universe''
(Grant No. XDB23000000) of the Chinese Academy of Sciences, the Natural Science Foundation
of Jiangsu Province (Grant No. BK20161096), and NASA NNX 13AH50G.


\providecommand{\href}[2]{#2}\begingroup\raggedright\endgroup

\end{document}